# A testable hypothesis on the possible role of electron spin in the origin of bio-homochirality

Wei Wang

CCMST, School of Physics, Harbin Institute of Technology, Harbin, China

Email: wwang_ol@hit.edu.cn

**Abstract**

   The emergence of biomolecular homochirality is a critical open question in the field of origins of life. In order to seek out an answer to this unsettled issue, a number of mechanisms have been offered over time, but it still remains a great challenge to date. In this paper, based on the hydrothermal vent theory for the origins of life, I tentatively put forward a new hypothesis that the prebiotic emergence of the uniform chirality of biomolecules might have been specifically determined by the spin state of electrons during their prebiotic syntheses on the surfaces of greigite, a mineral which has been frequently argued to have played an important role in the evolutionary context of life. An experimental model to test the hypothesis has also been proposed. Taken into consideration the possible widespread existence of greigite in submarine hydrothermal vent systems which have been frequently argued as a potential cradle for the origins of life, the suggested model, if could be experimentally demonstrated, may be suggestive of where and how life originated on early Earth.

## Background

Symmetry is a fundamental aspect of nature. For example, certain molecules exist in two forms which are symmetrical mirror images of each other. They are called chiral molecules. Common chemical synthesis generally produces equal amounts of the two forms of chiral molecules. In living systems, however, this symmetry is broken. Naturally occurring proteins are composed of L-amino acids but not D-forms, whereas DNA and RNA contain only D-sugars. How did this homochirality originally form in the context of the origin of life is a giant conundrum and has received much attention. Bada and Miller [1] claimed that the emergence of biomolecular handedness "must have occurred at the time of the origin of life or shortly thereafter". This and a few other similar arguments, grouped as "biogenic theories" [2], say that the self-propagating nature of life's chemistry inherently settles on one enantiomer in the racemic prebiotic world, with no need of any abiological source of asymmetry. But they did not explain why one specific enantiomer over the other was preferentially chosen. On the other hand, it has been frequently suggested that the development of the chiral homogeneities preceded the origin of life on early Earth, since heterochirality would inhibit the incorporation of monomers into biopolymer [2, 3], and distort the regular higher-order structures of biomacromolecules [2, 4] which are necessary preconditions for biological functions. This is called "abiotic mechanism" [2].

From the abiogenic point of view, a number of proposals have been offered to explain how the unique chirality of biomolecules could have arisen in the presumably racemic prebiotic world. Such hypothetical possibilities include asymmetric physical fields [2, 5-7], spontaneous symmetry breaking [8, 9], enantiomeric resolution via crystallization [10, 11], chiral enrichment by sublimation [12, 13], extraterrestrial delivery from outer space [5, 14], stereoselective adsorption [15], stereospecific polymerization [3, 16], dissymmetric decomposition [17-20] or synthesis [21] induced by circularly polarized light (CPL), and asymmetric autocatalysis [22, 23]. Here I have no intention to classify the various proposed pathways (for this, please see the review by Bonner [2]), or to discuss their merits or problems. But in general, most of these mechanisms take place by chance and/or by arrangement more than by nature, and also some experimental results are very limited in enantiomeric excess or to model compounds and reaction conditions [2, 24-26]. To date no conclusive scenario regarding the handedness issue has been described. The enduring puzzle of bio-homochirality still remains a challenge [2, 27].

When thinking about this question of much debate, I always entertain a notion that it should be linked to an origin-of-life theory because no chirality no function, and then no life. In this paper, I tentatively put forward a new hypothetic mechanism for the abiogenic emergence of bio-homochirality, based on the hydrothermal vent theory for the origins of life on early Earth, by using the synthesis of α-amino acids as a paradigm. Specifically speaking, the prebiotic uniform handedness of amino acids might result from the electron spin state during their abiotic synthesis on the surface of greigite, a mineral which has been frequently argued to have played an important role in the origins of life [28]. The spin-induced chirality selective rule in amino acid synthesis will firstly be elaborated. Subsequently, before setting out to talk about the electron spin property of greigite and its possible role in the asymmetric synthesis, I would give a general introduction to the hydrothermal vent theory and the possible widespread existence of greigite in the primitive hydrothermal vent systems. At last, an experimental model will be suggested for the validation of the hypothesis.

## Hypothesis

*Spin-induced chiral selectivity*

In living organisms, reversible amination of α-oxo acids is a group of reactions of key importance in amino acid metabolism. For instance, glutamate dehydrogenase and transaminase can regulate the homeostasis of certain α-amino acids and α-oxo acids [29]. A similar function in various bacteria is performed by a family of amino acid dehydrogenases [30, 31]. They catalyze the reversible transformation between almost all proteinic amino acids and their corresponding oxo acids [32],

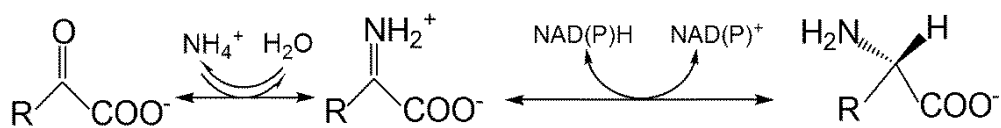

**Scheme 1**

In Scheme 1, the imine intermediate is reduced by the transfer of a hydride ion (a proton and two electrons) from coenzyme NAD(P)H to form L-amino acids. In this enzyme-catalyzed asymmetric reaction, it is generally accepted that the stereospecific regiostructure of the active site of the enzyme controls the handedness of the product. However, in the prebiotic world, how could a homochiral world of biomolecules have formed in the absence of enzymatic networks? A new "determinate mechanism"[2] I would propose here is that the enatioselective synthesis was determined by the spin state of the electrons in the reductive amination reaction.

Spin is an intrinsic property of all fermions whose spin quantum numbers ($m_s$) take half-integer values. For electrons, spin is divided into two well-entrenched camps, "spin up" (↑, $m_s = +1/2$) and "spin down" (↓, $m_s = -1/2$). In the absence of a magnetic field, these two states are present in statistically equal numbers and degenerate, just like the racemization of the two mirror images of a chiral molecule. In the presence of magnetic interactions, however, the two spin states can be split and well discriminated [33].

Electron spin is also determinate when filling electrons into atomic orbitals. For example, in the reductive amination reaction of an α-oxo acid, protons and electrons may attack the Schiff base separately or synergistically (Fig. 1). If synergistically, the reaction is referred to as a proton-coupled electron transfer (PCET) process [34]. The reduction of the C=N bond needs the transfer of two electrons, but always proceeds in two successive univalent steps, the intermediate state being a free radical [35, 36] (Fig. 1). The first negative electron is added to the positive iminium ion, producing a C radical intermediate. In the second step, there are four possible spin state combination forms between the radical intermediate and the second electron (or H atom) since each of them has two spin states (Fig. 1). According to the Pauli exclusion principle, however, two electrons in a bonding orbital can never have the same spin. Therefore, pathways (b) and (c) in Fig. 1 are forbidden. *The hypothesis is that pathways (a) and (d) might produce the two nonsuperposable mirror images of an amino acid, with different stereo-configurations from one another*. If channel (a) brought L-amino acids, channel (d) would yield D-amino acids. In other words, in a reaction between two radicals

(including electron and H atom), if the spin orientation of the unpaired electron of one radical is already defined, the stereo-configuration of the chiral product is determined as well. Here it should be noted that the spin state of the second electron is instantaneous. Once the new chemical bond forms, the electron will delocalize over the whole molecule and also its spin orientation will flip over and over again due to spin exchange interactions.

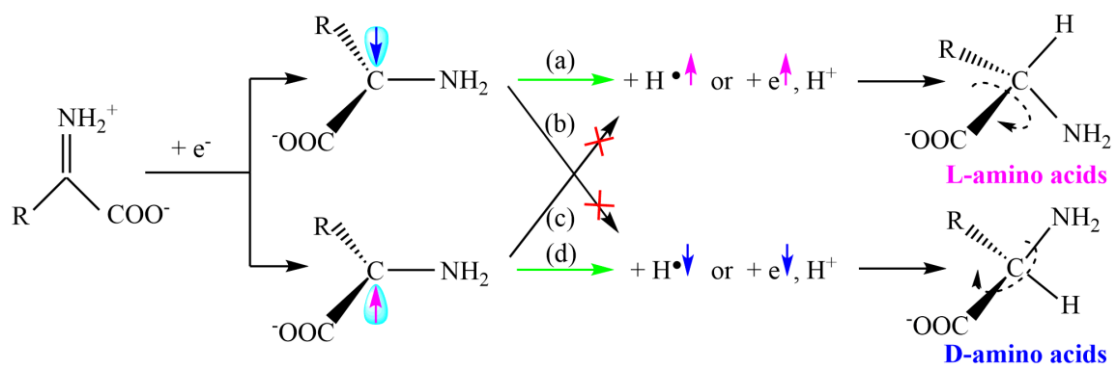

Fig. 1 Schematic illustration of the four possible spin state combinations in the reductive amination process of α-oxo acids and the hypothesis on the spin-polarized electron induced stereoselective synthesis of α-amino acids.

If the above conjecture was right, the chirality of the amino acid product would depend on the spin state of the second electron, or in a PCET reaction the second hydrogen atom. A neutral hydrogen atom consists of a proton and an electron (Fig. 2). The atomic hydrogen with the spins of the electron and the proton aligned in the same direction (parallel) has slightly more energy than one where the two spins are in opposite directions (antiparallel) [37]. This is an inherent quantum mechanical result, which arises from the hyperfine splitting of the 1$S$ ground state due to the interaction between the magnetic moments of the proton and electron. The splitting leads to two distinct energy levels separated by $\Delta E = 5.9 \times 10^{-6}$ eV. They represent two ground state forms of atomic hydrogen. When the relative spins change from parallel to antiparallel, a photon is emitted. This emission, famously known as the 21 cm line (corresponding to the energy of $5.9 \times 10^{-6}$ eV), is easily observed by radio telescopes [38]. However, it is unlikely to be detected in laboratory on Earth, because the transitions are rare and only in the interstellar medium are there enough atomic hydrogens and emission lines to be readily observed. When hydrogen atoms react with the radical intermediate in Fig. 1, as a result of magnetic dipole interactions, the antiparallel structure owning a lower energy might lead to the formation of L-amino acid which is also at a lower energy level than its corresponding D-amino acid [39, 40]. This may help explain why the natural enantiomers in terrestrial biochemistry always have intrinsically more stable low energies and prefer over its unnatural counterpart.

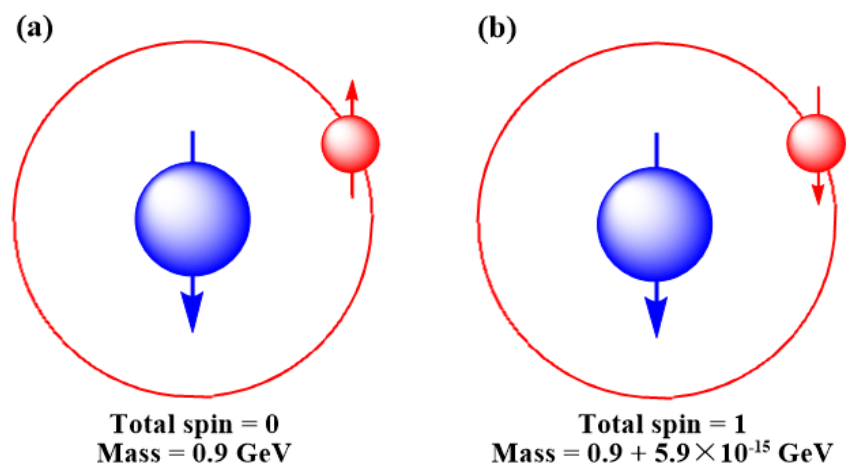

Fig. 2. In a hydrogen atom, aligning the spins of the proton (blue) and the electron (red) increases the atom's total spin from zero to one. If the electron revolves around the nucleus with antiparallel spin alignment (left), the system occurs in a lower energy state.

The above mechanism could also be applied to explain the asymmetric degradation of chiral molecules induced by CPL [18, 19, 41]. Photons of left- and right-CPL have opposite projections of angular momentum on their propagation direction. When a beam of CPL is absorbed by a molecule, polarized photons will exchange angular momenta with molecular electrons. In this way CPL can produce spin polarized electrons and vice versa [42, 43]. The resulted spin polarized electrons may induce the asymmetric decomposition of the molecule. Recently, a series of very beautiful studies contributed by Rosenberg et al. [44-46] and Dreiling and Gay [47] showed that spin-polarized electrons arising from magnetized Permalloy ($Fe_{0.2}Ni_{0.8}$), chiral self-assembled layer of DNA, and semiconductor GaAs all can trigger asymmetric dissociation of chiral molecules. Although the mechanism behind those observations is still open, a conclusion may be modestly drawn that there does exist a potential relationship between electron spin polarization and chiral discrimination. This would be called spin-induced chiral selectivity (SICS).

Based on the discussion above, the new hypotheses I would present in this paper looms clearer. The key content of the hypothesis deals with asymmetric synthesis but not degradation. The emergence of prebiotic homochirality might have its root in the SICS mechanism. The spin polarization constrains the symmetry of the wave functions involved with the radical coupling reaction (Fig. 1), resulting in a small energy difference between the two chiral forms of the product by which only one enantiomer is selected. This mechanism is based on a pure quantum mechanical effect, that is, the spin-spin or spin-orbit interaction. It is thanks to a source of electrons or H atoms with a single spin state. But where and how could such a homogenous electron spin state have formed in the prebiotic world? This will be discussed subsequently after an introduction to the hydrothermal vent theory for the origins of life on early Earth.

*Hydrothermal vent theory and greigite*

Hydrothermal vents are fissures in a planet's solid surface through which geothermally heated water issues [48]. At the bottom of an ocean, seawater infiltrates deeply down into the earth's crust through cracks. In the first type of vent, after heated by magmatic intrusions at oceanic spreading centers, the waters circulate and gush back into the ocean at temperatures up to 400 °C [49, 50]. Due to the high hydrostatic pressure (approaching to tens of megapascal) at the base of the convection cell, the waters may exist either in liquid form or as supercritical fluids, i.e., fluids at temperatures at or above their critical temperatures. Water-rock interactions in these conditions dissolve iron, manganese, zinc, and copper, along with $CO_2$, $H_2S$ and $CH_4$ out of the rock in what are, by now, acidic solutions. Hydrogen is also produced through the reduction of a portion of the circulating ocean water. On meeting present-day alkaline seawater calcium sulfate is precipitated, only to be rapidly replaced by iron, copper and zinc sulfide [48, 51]. On the early Earth when oceans were acidic, these metals would have been born directly into the ocean where they would have remained supersaturated until meeting alkaline waters [52, 53].

A second type of hydrothermal spring locates several kilometers away from the spreading zone. These are alkaline (pH 9-11) and cooler, attaining a temperature approaching 100 °C [54-56]. They are driven merely by the geothermal gradient and exothermic reactions [57]. Although negligible contents of the transition and post-transition metals are recorded in these fluids, they do also contain varying concentrations of $H_2$ and $CH_4$, and yet a small amount of formate and longer hydrocarbons [48, 58, 59]. The chimneys growing at these submarine springs today are also in marked contrast to the first type of vent. They comprise calcite, aragonite (both are $CaCO_3$) and brucite ($Mg(OH)_2$) [54, 60], though those growing in the Hadean likely consisted of ferrous/ferric oxyhydroxides, silica and sulfides [61, 62].

Since their first discovery in 1977 [63], the submarine hydrothermal vents have been frequently argued as a congenial site for the origins of terrestrial life [54, 64-69]. However, alkaline vents now seem more likely [56, 70-72]. Nevertheless, either scenario makes sense because many living organisms are found near the vents, especially a great diversity of extremophile archaea [73-75] which are generally believed to be the most primitive life forms on Earth [76].

In fact, these discoveries are only one of several reasons why in the last forty years interest has been focused on hydrothermal vents in general as potential sites for life's origin. Some other aspects are as follows: 1) the thermal fluids under high pressure are conducive to the prebiotic synthesis of organics [77, 78]. 2) The minerals at either type of vent could have acted as catalysts, templates, compartments, and power sources for not only the prebiotic synthesis of biomolecules [66, 69, 79-81], but also the crucial evolutionary nascence of ancient metabolic pathways and the emergence of the first cells/vesicles [70, 82-86]. 3) On early Earth, the submarine niche provided an ideal refuge where primitive life could have been protected against extensive meteorite impacts, hard UV, and partial vaporization of the ocean [78]. 4) In 2000, Rasmussen [87] reported a discovery of pyritic filaments in a 3,235-million-year-old deep-sea volcanogenic massive sulfide deposit from the Pilbara Craton of Australia. This fossil remains of thread-like microorganisms which are probably thermophilic chemotropic prokaryotes. 5) In all extant life forms, there exists a family of iron-sulfur proteins [88-90]. The Fe-S cluster structures at the active centers of these proteins are similar to some sulfide minerals. For instance, the [4Fe-4S] or [3Fe-4S] 'thiocubane' units in a class of ferredoxins look very like the cubic unit of greigite, $Fe_3S_4$ [70, 82, 86]. Therefore, a notion has been entertained that iron-sulfur proteins may evolve from those mineral structures sequestered by primordial abiotic peptides, and the whole later on to act as a biological electron transfer agent [90, 91]. All these concerns imply that the hydrothermal vents might be the initial hatchery for the last universal common ancestor of all extant life on Earth.

The original major sulfide precipitated on the mixing of the $HS^-$-bearing hydrothermal fluids and $Fe^{2+}$-bearing ocean water upon the chimneys would be nano-sized colloidal particles of amorphous FeS [66, 92]. With time, disordered FeS always reacts to form more stable iron sulfide phases such as metastable but well-organized tetragonal mackinawite ($FeS_m$) and cubic greigite, and ultimately ends up mainly as pyrite ($FeS_2$) [28, 93-96]. Mackinawite and greigite are rarely observed in today's marine sediments [94]. On the contrary, the most common sulfide mineral of sea-floor deposits from hydrothermal fumaroles is pyrite [97]. Nevertheless, a number of studies have implicated the mackinawite and magnetic greigite as essential intermediates for the formation of pyrite as follows: $Fe^{2+} + S^{2-} \rightarrow FeS \rightarrow Fe_3S_4 \rightarrow FeS_2$ [98-100]. Especially in the case of framboidal pyrite, it was suggested that the role of greigite as a precursor phase accounts for the formation of the unique raspberry-like shape of pyrite, by attributing its framboidal form to the aggregation of greigite microcrystals under magnetic forces and their subsequent pyritization [100]. Ohfuji and Akai [101] have examined pyrite framboids from various of sediments. They found that most of the microcrystals do have ferromagnetic greigite components in their central parts.

Chemically, the formation of $FeS_2$ from FeS requires the oxidation of S but not Fe, while in the FeS to $Fe_3S_4$ transition $Fe^{2+}$ is oxidized to $Fe^{3+}$, whereas $S^{2-}$ remains unchanged. Greigite is a mixed-valence compound, consisting of $Fe^{2+}$ and $Fe^{3+}$ centers in a 1:2 ratio. Thus it is clear that an oxidant is necessary for mackinawite to react to greigite, and also it is of key importance to tune the valence fluctuation of iron ions during the mineralization of greigite [102]. Indeed, weak inorganic or organic oxidants such as elemental sulfur [103-106], polysulfide anions [106, 107], an appropriate amount of oxygen [106, 108], and cysteine (possibly in the form of cystine) [106, 109-111] have assuredly been demonstrated to be able to affect the chemistry of iron sulfides and help the formation of greigite. White et al. [112] recently reported the small scale oxidation of freshly-precipitated mackinawite to greigite at temperatures of around 75 °C. They suggested a reaction mechanism whereby anoxic $H_2O$ is the oxidant, although this is still an open question.

Structurally, mackinawite has a layer structure, wherein an iron ion is linked in a tetragonal coordination to four equidistant sulfur ions (Fig. 3c). Greigite is a cubic inverse thiospinel mineral. The metal ions locate at two different interstices, tetrahedral A sites and octohedral B sites, between a close-packed cubic lattice of the sulfur ions (Fig. 3a). A sites are populated by $Fe^{3+}$ ions only and B sites by both $Fe^{3+}$ and $Fe^{2+}$ ions. Electron hopping between ions of the two different valences results in the large conductivity of greigite. Pyrite has a cubic crystal structure with $Fe^{2+}$ ions at the corners and face centers of the cube unit cell and dumbbell shaped disulfide $S_2^{2-}$ ions at the cube center and the midpoints of cube edges (Fig. 3b). As proposed earlier by Wang et al. [28], the transformation of mackinawite to greigite occurs through partial oxidation of the ferrous ions in the multilayer lattice followed by atom rearrangement. Whereas to pyrite, it needs complete lattice dissolution and disturbance [96]. An intuitive idea is that the former transition takes place more readily. However, since greigite is thermodynamically metastable with respect to pyrite, the anoxic oxidation of mackinawite in hydrothermal conditions always mainly produces pyrite [106], although in that process greigite maybe has existed as an intermediate [100].

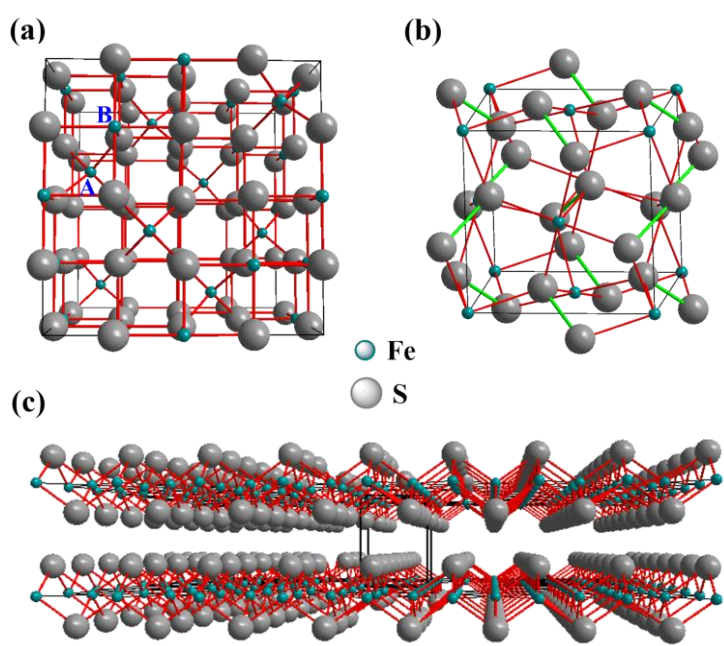

Fig. 3 Crystal structures of greigite (a), pyrite (b), and mackinawite (c). The black lines are just drawn to show the unit cell edges.

But interestingly, in the paper by Rickard and coworkers [96], they showed that trace amounts of aldehydes enable the formation of greigite with high purity from mackinawite and inhibit pyrite formation, whereas in the absence of aldehydes the products are almost entirely pyrite. This is because aldehydes suppress the lattice dissolution and damage at the surface sites of mackinawite. More recently, Wang et al. [28] demonstrated that α-oxo acids are more efficient than aldehydes in triggering the selective oxidation of FeS to greigite at 25-100 °C. They suggested that both α-oxo acids and aldehydes adsorb on the surfaces of mackinawite to form five-membered ring complexes, in which the Fe 3d electrons are pulled toward the carbonyl group of the organic molecules to produce pseudo $Fe^{3+}$ ions. The latter then drives the rearrangement of Fe atoms to form greigite, while the cubic close-packed S array retained in the new phase. These results confirm that a partially oxidized intermediate is necessary for the mackinawite-to-greigite transformation to occur. In addition, either mechanism implies that greigite rather than pyrite may have been produced from mackinawite in the primitive hydrothermal vents, since the above-mentioned carbonyl compounds could have been generated on the early Earth and then accumulated in the prebiotic ocean [113, 114]. Especially the α-oxo acids, that they are able to be *in situ* synthesized under submarine hydrothermal conditions [115] is apposite to the emergence of greigite in the hydrothermal deposits.

The pathway from mackinawite to greigite is suggestive of the origins of life in the hydrothermal vent systems, since the greigite crystal structure bears good affinity with the active centers of bacterial or archaeal enzymes [116] and is comparable to the cubanes comprising the active center of [4Fe-4S] enzymes [70]. More important is that the mineral greigite may serve as a spin filter to produce spin-polarized currents and then induce asymmetric syntheses. This will be discussed below in more detail.

*Electron spin properties of $Fe_3S_4$: implications for chiral synthesis*

Greigite ($Fe_3S_4$) is an iron thiospinel. It has the same crystal structure as magnetite ($Fe_3O_4$) and crystallizes in the inverse spinel structure. In physics, greigite is ferromagnetic, with the spin magnetic moments of the Fe cations in the tetrahedral sites oriented in the opposite direction to those in the octahedral sites, and a net magnetization. Both metal sites have high spin quantum numbers. Electron hopping is inferred to occur between high-spin ferric and ferrous ions in octahedral lattice positions, resulting in a metal-like conductivity of greigite. These special properties of greigite make it a half metal.

The half metal is an extreme case of completely spin polarized materials. It is metallic for one spin direction and insulating for the other [117]. In other words, a half-metal acts as a conductor to electrons of one spin orientation, but as an insulator to those of the opposite orientation (Fig. 4). Thus when a "racemic" current consisting of equal amounts of up-spin and down-spin electrons passes through a half metal like greigite, the latter can serve as a filter to produce a totally spin-polarized current.

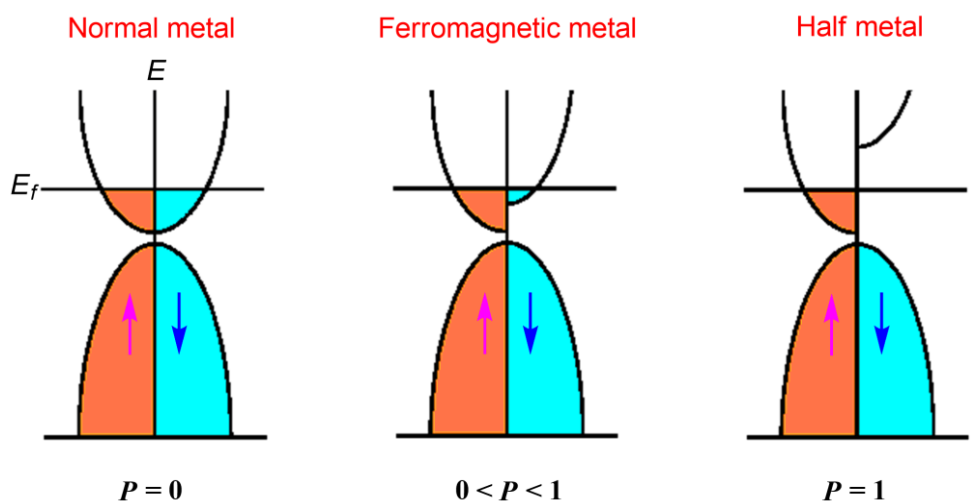

Fig. 4 Electronic band structures of the paramagnetic, ferromagnetic and half metal materials. Each side of the area marked with arrows indicates the densities of states of up-spin and down-spin electrons, respectively.

Spin polarization rate ($P$) is defined as the ratio of the density (D) of states of up-spin and down-spin electrons at a Fermi level ($E_f$), $P = (D_\uparrow - D_\downarrow)/(D_\uparrow + D_\downarrow)$. In normal metals or paramagnetic materials such as organic molecules, electrons all exist in pairs. The magnetic field of one electron is cancelled by an opposite magnetic field of the other electron in the pair. Thus the densities of states of up-spin and down-spin electrons at their Fermi levels are equal (if the two densities are both zero, it is called semiconductor or insulator), and their $P = 0$ (Fig. 4a). Some ferromagnetic materials such as Fe, Co, and Ni own unpaired 3d electrons and large spin-orbit coupling (SOC). Their $P$ values are larger than 0 but smaller than 1 because the densities of states of up-spin and down-spin electrons at $E_f$ are different (Fig. 4b). However, if a material has a band gap in the minority band but exhibits metallic behavior in the majority band, the density of state of the minority band is zero and as a result $P = 1$ (Fig. 4c). A half-meal like greigite is exactly such a material.

So the mineral greigite may act as a spin filter to produce polarized electrons with a uniform spin orientation and induce asymmetric synthesis through the SICS mechanism. If the assumption was true, greigite might have participated in the formation of the homochiral world in the prebiotic hydrothermal vents. However, the composition of an ancient hydrothermal sulfide chimney would have been heterogeneous. Some other nonmagnetic minerals such as pyrite, chalcopyrite ($FeCuS_2$), and sphalerite (ZnS) might have formed therein [48, 51, 118]. These nonmagnetic intergrowth species may do damage to the inherent spin density state of greigite. However, greigite is a magnetic mineral. It can be controlled by the primordial geomagnetic field or the natural remnant magnetism in rocks. This could have helped the rectification and directional accretion of greigite particles, and also the alignment of magnetic dipoles in the particles against thermal fluctuations [92, 119-121]. Binding to the field-controlled greigite particles would have caused a drastic reduction of phase space available to the reacting organics [120]. The team-up of greigite microstructures themselves, as well as along with organics, not only offers a mechanism for the framboid-like scaling-up of magnetic assemblies [100], but also permits the diffusing-in and diffusing-out of organic/inorganic "foods" and "wastes", respectively, just like a metabolic machine [120]. Meanwhile, this field-controlled system exerts regulation on electronic spin states and hence spin-selective chemical reactions [119] and molecular interactions [122]. When the natural electricity *in situ* generated between the separated hydrothermal fluids and seawater [80, 81, 123] is applied to the greigite electrode, a net spin polarized current is produced and then spin-selective chemical reactions happen to the adsorbed organics on the mineral's surfaces.

Later on, in the evolutionary context, $Fe_3S_4$ would have combined with prebiotic peptides and finally evolved into extant iron-sulfur proteins, whose Fe-S cluster active centers are strikingly similar to greigite in tiny structure [70]. During the prebiotic phase or the early stage after life arose, $Fe_3S_4$ and iron-sulfur clusters of the proto-proteins might have acted as spin filters to induce asymmetric synthesis of amino acids and other chiral biomolecules. Although still occur in extant ferredoxins, over time the Fe-S cluster structures and their catalytic function got partially pushed away from bioentities upon takeover by pure organic enzymes. For example, the amino acid dehydrogenases, which catalyze the asymmetric synthesis of amino acids (Scheme 1), comprise a cofactor of NAD(P)H but not Fe-S clusters. But this does not mean that the amino acid dehydrogenases in the early stage were definitely not a Fe-S cluster-dependent enzyme, since during evolution their iron sulfur clusters might have been replaced by NAD(P)H, as shown in the Entner-Doudoroff pathway of *P. furiosus* which belongs to the most primitive organisms [124]. Taking inspiration from this assumption, therefore, it might be expected that the NAD(P)H-coupled amino acid dehydrogenases in extant organisms might be evolutionary vestiges of a primordial metalloenzyme ancestor. In other words, in the prebiotic stage greigite participated in producing chiral amino acids through the SICS mechanism, then iron-sulfur proteins did this, and at last nonmetallic enzymes made of homochiral amino acids.

## Validation: a suggested test model

According to the above arguments, the half metal $Fe_3S_4$ can produce a net spin polarized electric current and drive asymmetric reduction for the synthesis of chiral chemicals.

To test the suggested hypothesis, one can prepare $Fe_3S_4$ electrodes to catalyze the asymmetric syntheses of chiral molecules such as α-amino acids (Scheme 1). When doing this, an external magnetic field is suggested to be added for spin state regulation. Thin films of Fe, Ni, or Fe/Ni alloy may also be considered as a candidate magnetic electrode because they are more easily prepared with high quality by using conventional electron beam evaporation [125] or magnetron sputtering [126] techniques, though their $P$ values are only about 0.3-0.5 [127].

## Discussions and conclusions

Scenarios to account for the origin of life deal with a series of cascading processes from inanimate chemicals to the first free-living cells. They look at basic chemistry of life—the molecular building blocks, the genetic code, the translation and transcription systems, the metabolic network and the cell membrane. They speak to how the universal attributes of life could have formed by retracing the path to the ancient times. Homochirality is a typical common feature of all extant life on Earth. Therefore, all theories for the origin of life must offer testable hypotheses to account for the source of the homochirality. If an abiotic mechanism for the primordial origin of chiral homogeneity is necessary, a scenario for the origin of life must be crosslinked to it.

In this paper, based on the hydrothermal vent theory for the origins of life, I tentatively put forward a new hypothesis that the original asymmetry of biomolecules might have emerged on the surfaces of mineral greigite. As a half metal, the mineral greigite might act as a spin filter to produce spin-polarized electrons, and then trigger the asymmetric synthesis of biomolecules. If we consider life as an "order", and the racemic prebiotic world as a "chaos", it is difficult to imagine how an "order out of chaos" could have taken place, because no chirality no life. However, since there are good evidences to suspect the widespread existence of greigite in primordial hydrothermal chimneys, the suggested reaction model, if could be experimentally demonstrated, may support the scenario that life originated in a local environment such as the hydrothermal vent systems on early Earth. The "organic order" of biomolecular chirality possibly arose out of the natural "inorganic order", electron spin. Life is highly-ordered. It has been frequently argued that life's emergence could be postulated as a continuum from Earth's inorganic geochemical processes to chemoautotrophic biochemical processes. Thus in the chemical evolutionary context of "spin → chiral biomolecules →…→ protobiont", physical rules such as the SICS mechanism might have provided a driving force for the origin of the secondary order — chirality, and the protolife was an "order out of order out of order". Life is just like a lone nocturnal boat floating on a boundless sea. We don't know where we came from. The suggested spin-induced chirality selective mechanism may draw a navigation line between the spin properties of inorganic mineral and the chiral features of organic life, and tell us the answer in which the spin and chirality are just like two light towers in the darkness.

**References**

[1] Bada JL, Miller SL. Racemization and the origin of optically-active organic-compounds in living organisms. Biosystems 1987, 20: 21-26.

[2] Bonner WA. The origin and amplification of biomolecular chirality. Orig Life Evol Biosph 1991, 2: 59-111.

[3] Joyce GF, Visser GM, van Boeckel CA, van Boom JH, Orgel LE, van Westrenen J. Chiral selection in poly(C)-directed synthesis of oligo(G). Nature 1984, 310: 602-604.

[4] Wang M, Zhou P, Wang J, Zhao Y, Ma H, Lu JR, Xu H. Left or right: How does amino acid chirality affect the handedness of nanostructures self-assembled from short amphiphilic peptides. J Am Chem Soc 2017, 139: 4185-4194.

[5] Bonner WA. Chirality and life. Orig Life Evol Biosph 1995, 25: 175-190.

[6] Feringa BL, van Delden RA. Absolute asymmetric synthesis: The origin, control, and amplification of chirality. Angew Chem Int Ed 1999, 38: 3418-3438.

[7] He YJ, Qi F, Qi SC. Effect of chiral helical force field on molecular helical enantiomers and possible origin of biomolecular homochirality. Med Hypotheses 1998, 51: 125-128.

[8] Frank FC. On spontaneous asymmetric synthesis. Biochim Biophys Acta 1953, 11: 459-463.

[9] Plasson R, Bersini H, Commeyras A. Recycling Frank: Spontaneous emergence of homochirality in noncatalytic systems. Proc Natl Acad Sci USA 2004, 101: 16733-16738.

[10] Shinitzky M, Nudelman F, Barda Y, Haimovitz R, Chen E, Deamer DW. Unexpected differences between D- and L-tyrosine lead to chiral enhancement in racemic mixtures - Dedicated to the memory of Prof. Shneior Lifson - A great liberal thinker. Orig Life Evol Biosph 2002, 32: 285-297.

[11] Viedma C. Chiral symmetry breaking during crystallization: Complete chiral purity induced by nonlinear autocatalysis and recycling. Phys Rev Lett 2005, 94: 065504

[12] Bellec A, Guillemin JC. A simple explanation of the enhancement or depletion of the enantiomeric excess in the partial sublimation of enantiomerically enriched amino acids. Chem Commun 2010, 46: 1482-1484.

[13] Fletcher SP, Jagt RB, Feringa BL. An astrophysically-relevant mechanism for amino acid enantiomer enrichment. Chem Commun 2007: 2578-2580.

[14] Elsila JE, Aponte JC, Blackmond DG, Burton AS, Dworkin JP, Glavin DP. Meteoritic amino acids: diversity in compositions reflects parent body histories. ACS Central Sci 2016, 2: 370-379.

[15] Hazen RM, Mineral surfaces and the prebiotic selection and organization of biomolecules. Am Mineralog 2006, 91: 1715-1729.

[16] Sandars PG. A toy model for the generation of homochirality during polymerization. Orig Life Evol Biosph 2003, 33: 575-587.

[17] Meierhenrich UJ, Nahon L, Alcaraz C, Bredehoft JH, Hoffmann SV, Barbier B, Brack A. Asymmetric vacuum UV photolysis of the amino acid leucine in the solid state. Angew Chem Int Ed 2005 44: 5630-5634.

[18] Shimizu Y, Kawanishi S. An efficient enantiomeric enrichment of tartaric acid using a highly intense circularly polarized light. Chem Commun 1996: 819-820.

[19] Noorduin WL, Bode AA, van der Meijden M, Meekes H, van Etteger AF, van Enckevort WJ, Christianen PC, Kaptein B, Kellogg RM, Rasing T, Vlieg E. Complete chiral symmetry breaking of an

amino acid derivative directed by circularly polarized light. Nat Chem 2009, 1: 729-732.

[20] Bailey J, Chrysostomou A, Hough JH, Gledhill TM, McCall A, Clark S, Menard F, Tamura M. Circular polarization in star-formation regions: Implications for biomolecular homochirality. Science 1998, 281: 672-674.

[21] de Marcellus P, Meinert C, Nuevo M, Filippi J, Danger G, Deboffle D, Nahon L, d'Hendecourt LLS, Meierhenrich UJ. Non-racemic amino acid production by ultraviolet irradiation of achiral interstellar ice analogus with circularly polarized light. Astrophys J Lett 2011, 727: L27.

[22] Soai K, Shibata T, Morioka H, Choji K. Asymmetric autocatalysis and amplification of enantiomeric excess of a chiral molecule. Nature 1995, 378: 767-768.

[23] Blackmond DG. Asymmetric autocatalysis and its implications for the origin of homochirality. Proc Natl Acad Sci USA 2004, 101: 5732-5736.

[24] Fajszi C, Czégé J. Amplification of asymmetry in polymerization. J Theor Biol 1981, 88: 523-531.

[25] Blackmond DG. Challenging the concept of "recycling" as a mechanism for the evolution of homochirality in chemical reactions. Chirality 2009, 21: 359-362.

[26] Evgenii K, Wolfram T. The role of quartz in the origin of optical activity on earth. Orig Life Evol Biosph 2000, 30: 431-434.

[27] Jones N. Frontier experiments: Tough science. Nature 2012, 481: 14-17.

[28] Wang W, Song Y, Wang X, Yang Y, Liu X. Alpha-oxo acids assisted transformation of FeS to $Fe_3S_4$ at low temperature: Implications for abiotic, biotic, and prebiotic mineralization. Astrobiology 2015, 15: 1043-1051.

[29] Berg JM, Tymoczko JL, Stryer L. *Biochemistry*. 2002, New York: WH Freeman and Company.

[30] Nitta Y, Yasuda Y, Tochikubo K, Hachisuka Y. L-amino acid dehydrogenases in Bacillus subtilis spores. J Bacteriol 1974, 117: 588-592.

[31] Oikawa, T., et al., *Psychrophilic valine dehydrogenase of the antarctic psychrophile, Cytophaga sp. KUC-1: purification, molecular characterization and expression.* Eur J Biochem, 2001. **268**(16): p. 4375-83.

[32] Mohammadi HS, Omidinia E, Lotfi AS, Saghiri R. Preliminary report of $NAD^+$-dependent amino acid dehydrogenase producing bacteria isolated from soil. Iran Biomed J 2007, 11: 131-135.

[33] Phipps TE, Taylor JB, The magnetic moment of the hydrogen atom. Phys Rev 1927, 29: 309-320.

[34] Reece SY, Nocera DG. Proton-coupled electron transfer in biology: results from synergistic studies in natural and model systems. Annu Rev Biochem 2009, 78: 673-699.

[35] Commoner B, Lippincott BB, Passonneau JV. Electron-spin resonance studies of free-radial intermediates in oxidation-reduction enzyme systems. Proc Natl Acad Sci USA 1958, 44: 1099-1110.

[36] Kim J, Darley DJ, Buckel W, Pierik AJ. An allylic ketyl radical intermediate in clostridial amino-acid fermentation. Nature 2008, 452: 239-242.

[37] Rith K, Schafer A. The mystery of nucleon spin. Sci Am 1999, 281: 58-63.

[38] Furlanetto SR, Oh SP, Briggs FH. Cosmology at low frequencies: The 21 cm transition and the high-redshift Universe. Phys Rep 2006, 433: 181-301.

[39] Macdermott AJ. Electroweak enantioselection and the origin of life. Orig Life Evol Biosph 1995, 25: 191-199.

[40] Tranter GE. Parity violation and the origins of biomolecular handedness. Biosystems 1987, 20: 37-48.

[41] Balavoine G, Moradpour A, Kagan HB. Preparation of chiral compounds with high optical purity by irradiation with circularly polarized light, a model reaction for the prebiotic generation of optical

activity. J Am Chem Soc 1974, 96: 5152-5158.

[42] Dyakonov MI. Spin physics in semiconductors. 2008, Springer-Verlag Berlin Heidelberg.

[43] Nishizawa N, Nishibayashi K, Munekata H. Pure circular polarization electroluminescence at room temperature with spin-polarized light-emitting diodes. Proc Natl Acad Sci USA 2017, 114: 1783-1788.

[44] Rosenberg RA. Spin-polarized electron induced asymmetric reactions in chiral molecules. Top Curr Chem 2011, 298: 279-306.

[45] Rosenberg RA, Abu Haija M, Ryan PJ. Chiral-selective chemistry induced by spin-polarized secondary electrons from a magnetic substrate. Phys Rev Lett 2008, 101: 178301.

[46]Rosenberg RA, Mishra D, Naaman R. Chiral selective chemistry induced by natural selection of spin-polarized electrons. Angew Chem Int Ed 2015, 54: 7295-7298.

[47] Dreiling JM, Gay TJ. Chirally sensitive electron-induced molecular breakup and the Vester-Ulbricht hypothesis. Phys Rev Lett 2014, 113: 118103.

[48] Tivey MK. Generation of seafloor hydrothermal vent fluids and associated mineral deposits. Oceanography 2007, 20: 50-65.

[49] Douville E, Vharlou JL, Oelkers EH, Bienvenu P, Jove Colon CF, Donval JP, Fouquet Y, Prieur D, Appriou P. The rainbow vent fluids (36 degrees 14'N, MAR): the influence of ultramafic rocks and phase separation on trace metal content in Mid-Atlantic Ridge hydrothermal fluids. Chem Geol 2002, 184: 37-48.

[50] von Damn KL. Seafloor hydrothermal activity: Black smoker chemistry and chimneys. Annu Rev Earth Planet Sci 1990, 18: 173-204.

[51] Haymon RM, Kastner M. Hot spring deposits on the East Pacific Rise at 21°N: preliminary description of mineralogy and genesis. Earth Planet Sci Lett 1981, 53: 363-381.

[52] Macleod G, McKeown C, Hall AJ, Russell M. Hydrothermal and oceanic pH conditions of possible relevance to the origin of life. Orig Life Evol Biosph 1994, 24: 19-41.

[53] Fitzsimmons JN, Boyle EA, Jenkins WJ. Distal transport of dissolved hydrothermal iron in the deep south Pacific ocean. Proc Nat Acad Sci USA 2014, 111: 16654-16661.

[54] Kelley DS, Karson JA, Blackman DK, Früh-Green GL, Butterfield DA, Lilley MD, Olson EJ, Schrenk MO, Roe KK, Lebon GT, Rivizzigno P, and the AT3-60 Shipboard Party. An off-axis hydrothermal vent field near the Mid-Atlantic Ridge at 30 degrees N. Nature 2001, 412: 145-149.

[55] Früh-Green GL, Kelley DS, Bernasconi SM, Karson JA, Ludwig KA, Butterfield DA, Boschi C, Proskurowski G. 30,000 years of hydrothermal activity at the Lost City vent field. Science 2003, 301: 495-498.

[56] Martin W, Baross J, Kelley D, Russell MJ. Hydrothermal vents and the origin of life. Nat Rev Microbiol 2008, 6: 805-814.

[57] Lowell RP, Rona PA. Seafloor hydrothermal systems driven by the serpentinization of peridotite. Geophys Res Lett 2002, 29: 1531.

[58] Proskurowski G, Lilley MD, Seewald JS, Früh-Green GL, Olson EJ, Lupton JE, Sylva SP, Kelley DS. Abiogenic hydrocarbon production at Lost City hydrothermal field. Science 2008, 319: 604-607.

[59] Lang SQ, Butterfield DA, Schulte M, Kelley DS, Lilleya MD. Elevated concentrations of formate, acetate and dissolved organic carbon found at the Lost City hydrothermal field. Geochim Cosmochim Acta 2010, 74: 941-952.

[60] Ludwig KA, Kelley DS, Butterfield DA, Nelson BK, Früh-Green G. Formation and evolution of carbonate chimneys at the Lost City Hydrothermal Field. Geochim Cosmochim Acta 2006, 70: 3625-3645.

[61] Mielke RE, Russell MJ, Wilson PR, McGlynn SE, Coleman M, Kidd R, Kanik I. Design, fabrication, and test of a hydrothermal reactor for origin-of-life experiments. Astrobiology 2010, 10: 799-810.

[62] Mielke RE, Robinson KJ, White LM, McGlynn SE, McEachern K, Bhartia R, Kanik I, Russell MJ. Iron-sulfide-bearing chimneys as potential catalytic energy traps at life's emergence. Astrobiology 2011, 11: 933-950.

[63] Corliss JB, Dymond J, Gordon LI, Edmond JM, von Herzen RP, Ballard RD, Green K, Williams D, Bainbridge A, Crane K, van Andel TH. Submarine thermal springs on the Galapagos Rift. Science 1979, 203: 1073-1083.

[64] Corliss JB, Baross JA, Hoffman SE. An hypothesis concerning the relationships between submarine hot springs and the origin of life on Earth. Oceanolog Acta 1981, Special issue: 59-69.

[65] Baross JA, Hoffman SE. Submarine hydrothermal vents and associated gradient environments as sites for the origin and evolution of life. Orig Life 1985, 15: 327-345.

[66] Wächtershäuser G. Before enzymes and templates: Theory of surface metabolism. Microbiol Rev 1988, 52: 452-484.

[67] Holm NG. Why are hydrothermal systems proposed as plausible environments for the origin of life. Orig Life Evol Biosph 1992, 22: 5-14.

[68] Russell MJ, Nitschke W, Branscomb E. The inevitable journey to being. Philosoph Transact Royal Soc B 2013, 368: 20120254.

[69] Russell MJ, Hall AJ, Cairns-Smith AG, Braterman PS. Submarine Hot Springs and the Origin of Life. Nature 1988, 336: 117.

[70] Russell MJ, Hall AJ. The emergence of life from iron monosulphide bubbles at a submarine hydrothermal redox and pH front. J Geolog Soc 1997, 154: 377-402.

[71] Martin W, Russell MJ. On the origin of biochemistry at an alkaline hydrothermal vent. Philosoph Transact Royal Soc B 2007, 362: 1887-1925.

[72] Russell MJ. The importance of being alkaline. Science 2003, 302: 580-581.

[73] Moyer CL, Tiedje JM, Dobbs FC, Karl DM. Diversity of deep-sea hydrothermal vent Archaea from Loihi seamount, Hawaii. Deep Sea Res Part II: Top Stud Oceanogr 1998, 45: 303-317.

[74] Takai K, Horikoshi K. Genetic diversity of archaea in deep-sea hydrothermal vent environments. Genetics 1999, 152: 1285-1297.

[75] Raguenes G, Meunier J, Antoine E, Godfroy A, Caprais J, Lesongeur F, Guezennec J, Barbier G. Biodiversity of hyperthermophilic Archaea from hydrothermal vents of the East Pacific Rise. Comptes Rendus De L Academie Des Sciences Serie III 1995, **318**: 395-402.

[76] Woese CR, Kandler O, Wheelis ML. Towards a natural system of organisms: proposal for the domains Archaea, Bacteria, and Eucarya. Proc Nat Acad Sci USA 1990, 87: 4576-4579.

[77] Shock EL, Schulte MD. Organic synthesis during fluid mixing in hydrothermal systems. J Geophys Res-Planets 1998, 103: 28513-28527.

[78] Holm NG. Why Are Hydrothermal Systems Proposed as Plausible Environments for the Origin of Life. Orig Life Evol Biosph 1992, 22: 5-14.

[79] Hazen RM. Life's rocky start. Sci Am 2001, 284: 76-85.

[80] Barge LM, Abedian Y, Russell MJ, Doloboff IJ, Cartwright JH, Kidd RD, Kanik I. From chemical gardens to fuel cells: Generation of electrical potential and current across self-assembling iron mineral membranes. Angew Chem Int Ed 2015, 54: 8184-8187.

[81] Yamamoto M, Nakamura R, Oguri K, Kawagucci S, Suzuki K, Hashimoto K, Takai K. Generation of electricity and illumination by an environmental fuel cell in deep-sea hydrothermal vents. Angew

Chem Int Ed 2013, 52: 10758-10761.

[82] Russell MJ, Martin W. The rocky roots of the acetyl-CoA pathway. Trends Biochem Sci 2004, 29: 358-363.

[83] McGlynn SE, Kanik I, Russell MJ. Peptide and RNA contributions to iron-sulphur chemical gardens as life's first inorganic compartments, catalysts, capacitors and condensers. Philosoph Trans Royal Soc A 2012, 370: 3007-3022.

[84] Koonin EV, Martin W. On the origin of genomes and cells within inorganic compartments. Trends Genet 2005, 21: 647-654.

[85] Wächtershäuser G. On the chemistry and evolution of the pioneer organism. Chem Biodivers 2007, **4**: 584-602.

[86] Wang W, Yang B, Qu Y, Liu X, Su W. $FeS/S/FeS_2$ redox system and its oxidoreductase-like chemistry in the Iron-Sulfur world. Astrobiology 2011, 11: 471-476.

[87] Rasmussen B. Filamentous microfossils in a 3,235-million-year-old volcanogenic massive sulphide deposit. Nature 2000, 405: 676-679.

[88] Beinert H, Holm RH, Munck E. Iron-sulfur clusters: Nature's modular, multipurpose structures. Science 1997, 277: 653-659.

[89] Lill R. Function and biogenesis of iron-sulphur proteins. Nature 2009, 460: 831-838.

[90] Hall DO, Cammack R, Rao KK. Role for ferredoxins in the origin of life and biological evolution. Nature 1971, 233: 136-138.

[91] Russell MJ, Hall AJ, Boyce AJ, Fallick AE. On hydrothermal convection systems and the emergence of life. Econ Geol 2005, 100: 419-438.

[92] Mitra-Delmotte G, Mitra AN. Magnetism, entropy, and the first nano-machines. Central Europ J Phys 2010, 8: 259-272.

[93] Wolthers M, Charlet L, van Der Linde PR, Rickard D, van Der Weijden CH. Surface chemistry of disordered mackinawite (FeS). Geochim Cosmochim Acta 2005, 69: 3469-3481.

[94] Ohfuji H, Rickard D. High resolution transmission electron microscopic study of synthetic nanocrystalline mackinawite. Earth Planet Sci Lett 2006, 241: 227-233.

[95] Rickard D, Luther GW. Chemistry of iron sulfides. Chem Rev 2007, 107: 514-562.

[96] Rickard D, Butler IB, Oldroyd A. A novel iron sulphide mineral switch and its implications for Earth and planetary science. Earth Planet Sci Lett 2001, 189: 85-91.

[97] Ikehata K, Suzuki R, Shimada K, Ishibashi J, Urabe T. Mineralogical and geochemical characteristics of hydrothermal minerals collected from hydrothermal vent fields in the southern Mariana spreading center, in Subseafloor Biosphere Linked to Hydrothermal Systems. Ishibashi J, Okino K, Sunamura M (Editors). 2015, Springer Japan. p275-287.

[98] Sweeney R, Kaplan IR. Pyrite framboid formation: Laboratory synthesis and marine sediments. Econ. Geol 1973, 68: 618-634.

[99] Schoonen MAA, Barnes HL, Reactions forming pyrite: Nucleation of $FeS_2$ below 100 °C. Geochim Cosmochim Acta 1991, 55:1495-1504.

[100] Wilkin RT, Barnes HL. Formation processes of framboidal pyrite. Geochim Cosmochim Acta 1997, 61: 323-339.

[101] Ohfuji H, Akai J. Icosahedral somain structure of framboidal pyrite. Am Mineralog 2002, 87: 176-180.

[102] Liao TQ, Wang W, Song Y, Wang X, Yang Y, Liu X. HMTA-assisted one-pot synthesis of greigite nano-platelet and its magnetic properties. J Mater Sci Technol 2015, 31: 895-900.

[103] Beal JHL, Prabakar S, Gaston N, The GB, Etchegoin PG, Williams G, Tilley RD. Synthesis and comparison of the magnetic properties of iron sulfide spinel and iron oxide spinel nanocrystals. Chem Mater 2011, 23: 2514-2517.

[104] Paolella A, George C, Povia M, Zhang Y, Krahne R, Gich M, Genovese A, Falqui A, Longobardi M, Guardia P, Pellegrino T, Manna L. Charge transport and electrochemical properties of colloidal greigite ($Fe_3S_4$) nanoplatelets. Chem Mater 2011, 23: 3762-3768.

[105] Dekkers MJ, Schoonen MAA. Magnetic properties of hydrothermally synthesized greigite ($Fe_3S_4$). 1. Rock magnetic parameters at room temperature. Geophys J Int 1996, 126: 360-368.

[106] Wilkin RT, Barnes HL. Pyrite formation by reactions of iron monosulfides with dissolved inorganic and organic sulfur species. Geochim Cosmochim Acta, 1996, 60: 4167-4179.

[107] Wada H. The synthesis of greigite from a polysulfide solution at about 100 °C. Bullet Chem Soc Jap 1977, 50: 2615-2617.

[108] Vasilenko IV, Cador O, Ouahab L, Pavlishchuk VV. Effect of the production conditions on the size and magnetic characteristics of iron sulfide $Fe_3S_4$ nanoparticles. Theor Experim Chem 2010, 46: 322-327.

[109] Liu XG, Feng C, Bi N, Sun Y, Fan J, Lv Y, Jin C, Wang Y, Li. Synthesis and electromagnetic properties of $Fe_3S_4$ nanoparticles. Ceram Int 2014, 40: 9917-9922.

[110] Cao F, Hu W, Zhou L, Shi WD, Song SY, Lei YQ, Wang S, Zhang HJ. 3D $Fe_3S_4$ flower-like microspheres: high-yield synthesis via a biomolecule-assisted solution approach, their electrical, magnetic and electrochemical hydrogen storage properties. Dalton Transact 2009: 9246-9252.

[111] He Z, Yu S, Zhou X, Li X, Qu J. Magnetic-field-induced phase-selective synthesis of ferrosulfide microrods by a hydrothermal process: Microstructure control and magnetic properties. Adv Function Mater 2006, 16: 1105-1111.

[112] White LM, Bhartia R, Stucky GD, Kanik I, Russell M. Mackinawite and greigite in ancient alkaline hydrothermal chimneys: Identifying potential key catalysts for emergent life. Earth Planet Sci Lett 2015, 430: 105-114.

[113] Pinto JP, Gladstone GR, Yung YL. Photochemical production of formaldehyde in Earth's primitive atmosphere. Science 1980, 210: 183-185.

[114] Pizzarello S, Weber AL. Prebiotic amino acids as asymmetric catalysts. Science 2004, 303: 1151.

[115] Cody GD, Boctor NZ, Filley TR, Hazen RM, Scott JH, Sharma A, Yoder HS Jr. Primordial carbonylated iron-sulfur compounds and the synthesis of pyruvate. Science 2000, 289: 1337-1340.

[116] Nitschke W, McGlynn SE, Milner-White EJ, Russell MJ. On the antiquity of metalloenzymes and their substrates in bioenergetics. Biochim Biophys Acta-Bioenerget 2013, 1827: 871-881.

[117] Wang XL, Dou SX, Zhang C. Zero-gap materials for future spintronics, electronics and optics. NPG Asia Mater 2010, 2: 31-38.

[118] Nakamura R, Takashima T, Kato S, Takai K, Yamamoto M, Hashimoto K. Electrical current generation across a black smoker chimney. Angew Chem Int Ed 2010, 49: 7692-7694.

[119] Mitra-Delmotte G, Mitra AN. Magnetism, FeS colloids, and origins of life, in The legacy of Alladi Ramakrishnan in the mathematical sciences. Alladi K, Klauder JR, Rao CR (Editors). 2010 Springer, p 529-564.

[120] Mitra-Delmotte G, Mitra AN. Field-control, phase-transitions, and life's emergence. Front Physiol, 2012, 3: 366.

[121] Mitra-Delmotte G, Mitra AN. Softening the "crystal scaffold" for life's emergence. Phys Res Inter 2012: 232864.

[122] Kumar A, Capua E, Kesharwani MK, Martin JML, Sitbon E, Waldeck DH, NaamanR. Chirality-induced spin polarization places symmetry constraints on biomolecular interactions. Proc Natl Acad Sci USA 2017, 114: 2474-2478.

[123] Herschy B, Whicher A, Camprubi E, Watson C, Dartnell L, Ward J, Evans JR, Lane N. An origin-of-life reactor to simulate alkaline hydrothermal vents. J Mol Evol 2014, 79: 213-227.

[124] Daniel RM, Danson MJ. Did primitive microorganisms use nonhem iron proteins in place of NAD/P. J Mol Evol 1995, 40: 559-563.

[125] Miyazaki T, Ajima T, Sato F. Dependence of magnetoresistance on thickness and substrate temperature for 82Ni-Fe alloy film. J Magnet Magnet Mater 1989, 81: 86-90.

[126] Nagura H, Saito K, Takanashi K, Fujimori H. Influence of third elements on the anisotropic magnetoresistance in permalloy films. J Magnet Magnet Mater 2000, 212: 53-58.

[127] Moodera JS, Mathon G. Spin polarized tunneling in ferromagnetic junctions. J Magnet Magnet Mater 1999, 200: 248-273.